\documentclass[11pt,a4paper]{article}
\usepackage{amsmath,amsfonts,amsthm}
\usepackage{mathrsfs}
\usepackage{mathptmx}
\usepackage[hmargin=1in,vmargin=1in]{geometry}
\usepackage{url}
\usepackage[authoryear]{natbib}
\usepackage{url}
\usepackage{color}
\usepackage{tikz}
\usepackage{lineno}
\usepackage{subcaption}
\usepackage[markers,figuresonly,nolists]{endfloat}

\newtheorem{thm}{Theorem}

\theoremstyle{remark}
\newtheorem{rem}[thm]{Remark}

\theoremstyle{definition}

\renewcommand{\d}{\mathrm{d}}
\newcommand{\e}{\mathrm{e}}
\newcommand{\R}{\mathbb R}

\begin{document}

\title{\Large \bf  A spatio-temporal analogue of the Omori-Utsu law of aftershock sequences}
\author{Marianito R. Rodrigo\thanks{School of Mathematics and Applied Statistics, University of Wollongong, Wollongong, New South Wales, Australia.
E-mail:~{\tt marianito\_rodrigo@uow.edu.au}}
}
\date{\today}
\maketitle

\begin{abstract}
A spatio-temporal version of the well-known Omori-Utsu law of aftershock sequences is proposed. This `diffusive Omori-Utsu law' satisfies a nonlinear partial differential equation~(PDE). A similarity reduction is obtained that reduces the PDE to an ordinary differential equation~(ODE). A nonzero constant solution of this ODE leads to the usual Omori-Utsu law. An exact and explicit similarity solution is found that corresponds to the original Omori law. An initial value problem for the `diffusive Omori-Utsu law' is also considered, and whose spatio-temporal dynamics are described by bounding functions that satisfy nonlinear, but linearisable, PDEs. Numerical results are also provided.

 
\end{abstract}


\section{Introduction}

An aftershock is a sudden movement of the earth's surface that typically follows an earthquake and is less violent than the first main shock. Aftershocks are characterised by their clustering properties in both time and space.

The temporal clustering of aftershocks is described by the well-known Omori law~\citep{Om1894a,Om1894b}
\begin{equation}
\label{omori}
n(t) = \frac{K}{t + c},
\end{equation}
where $n(t)$ is the frequency of aftershocks per unit time interval at time~$t$, and $K > 0$ and $c \ge 0$ are constants. \citet{Ut1957} emphasised that the decay of aftershock activity of several earthquakes is somewhat faster than that predicted by the Omori law~\eqref{omori} and proposed the model
\begin{equation}
\label{utsu}
n(t) = \frac{K}{(t + c)^p},
\end{equation}
where $p > 0$ (typically ranges from $p = 0.9$ to $p = 1.5$). Formula~\eqref{utsu} is nowadays referred to as the Omori-Utsu law. The power law reveals the `long-lived' nature of aftershock activity, in constrast to the exponential law arising in most decay laws in physics~\citep{UtOgMa1995}.

To date, other models have been put forward to generalise the Omori-Utsu law. For example, \citet{Ot1985,Ot1987} suggested a compound model
\begin{equation}
\label{otsuka}
n(t) = \frac{K \e^{-\alpha t}}{(t + c)^p},
\end{equation}
where $\alpha \ge 0$, which recovers \eqref{utsu} when $\alpha = 0$. For large~$t$, the effect of the exponential function predominates. This exponential decay of activity in later periods has been used for some aftershock sequences~\citep{UtOgMa1995}.

\citet{Mi2015}, using three different declustering techniques, showed that all aftershock sequences tested in three regional earthquake catalogs in California and Taiwan followed a stretched exponential law instead of a power law.

\cite{SaVe2019} employed a fractional-order integral equation to derive a double power law model, and the Omori-Utsu and stretched exponential models are obtained as limiting cases.

\citet{LoGa2006} evaluated the efficiency of various models in describing the time decay of aftershock rates for earthquakes in California and Italy. They showed that the Omori-Utsu law~\eqref{utsu} with $c = 0$ had a better performance in terms of modelling and forecasting. Moreover, they concluded that the paramater~$c$ does not represent a general property of aftershock sequences in California and Italy but its presence is likely induced in most cases by catalog incompleteness in the first times after the main shock. The difficulties of estimating the true value of $c$ for general aftershock sequences were pointed out in the survey paper by \citet[p.~11]{UtOgMa1995}.

It is fairly established that the Omori-Utsu law is purely empirical and \eqref{utsu}, or its particular case~\eqref{omori}, is a fitting formula. In an attempt to show that \eqref{omori} is in fact a fundamental law, \citet{Gu2016,Gu2017} proposed a novel interpretation of the Omori law by drawing an analogy with the decrease in the ionospheric plasma density as a result of recombination of charges of opposite sign. If $n_+(t)$ (respectively, $n_-(t)$) is the density of positive (respectively, negative) charges at time~$t$, and $n = \frac{1}{2} (n_+ + n_-)$, then the equation of recombination is
$$
\frac{\d n}{\d t} = -\sigma n_+ n_-,
$$ 
where $\sigma$ is the (positive) coefficient of recombination. As $n_+ \simeq n_-$ with a high degree of accuracy in the ionosphere, there holds approximately
\begin{equation}
\label{density-ode}
\frac{\d n}{\d t} = -\sigma n^2. 
\end{equation}
To complete the analogy, \citet{Gu2016,Gu2017} argued that if we now let $n(t)$ represent the number of active faults in the epicentral region of the main shock (as opposed to the frequency of aftershocks per unit time interval, the usual interpretation) at time~$t$, then $n(t)$ would also satisfy \eqref{density-ode} but with $\sigma$ representing the mean frequency of deactivation of the active faults. It is straightforward to show that the Omori law~\eqref{omori} satisfies \eqref{density-ode} with $\sigma = \frac{1}{K}$ and the initial condition is $n(0) = \frac{K}{c}$. Note that the above analogy is not necessarily valid for the Omori-Utsu law~\eqref{utsu}. 

\citet{Fa2020} gave Lagrangian and Hamiltonian formulations of the Omori law and provided an interesting analogy with the cosmic Big Rip. The foreshock phase corresponds to an accelerating universe approaching the Big Rip, the main shock to the Big Rip singularity and the aftershock to a contracting universe.

The spatial clustering properties of aftershocks is more complex and generally less well understood~\citep{HeOuSo2003}. Several studies have reported the presence of `aftershock diffusion', i.e. the phenomenon of expansion or migration of the aftershock zone with time~\citep{Im1981,TaKa1985a,TaKa1985b}. After the main shock occurs, most aftershocks are located close to the rupture plane of the main shock and then migrate away at a velocity ranging from $1 \, \mathrm{km}/\mathrm{h}$ to $1 \, \mathrm{km}/\mathrm{yr}$~\citep{RySa2001}. Aftershock diffusion is not universally observed~\citep{Sh1993} but is important in some areas than in others~\citep{TaKa1985a}. 

Another interpretation of aftershock diffusion comes from physical arguments. Aftershock diffusion is usually interpreted as a diffusion of the stress induced by the main shock, e.g.~by a viscous relaxation process~\citep{RySa2001} or by a fluid transfer in the crust~\citep{NuBo1972,BoNu2002}. However, a stress diffusion process is not necessary to explain aftershock diffusion~\citep{HeOuSo2003}.

Multiple triggering is an alternative interpretation of aftershock diffusion that does not rely on any stress diffusion process. The concept of triggered seismicity has been quantitatively investigated in \citet{HeSo2002a,HeSo2002b,HeSo2003} in the context of the epidemic-type aftershock~(ETAS) model of earthquakes viewed as pointwise events. The ETAS~model was introduced in \citet{Og1988} and in a different form in \citet{KaKn1980}. 

\citet{HeOuSo2003} analysed aftershock sequences in California to test for evidence of spatio-temporal diffusion. They proposed a mechanism based on multiple cascades of triggering to explain aftershock diffusion. Using two methods, they confirmed that diffusion of seismic activity is very weak, in fact much weaker than previously reported.

In this article, following the interpretation of the Omori law proposed by \cite{Gu2016}, we consider a spatio-temporal generalisation of the Omori law~\eqref{omori}\footnote{In fact, we will use the more general compound model~\eqref{otsuka} as our starting point.} by studying the nonlinear partial differential equation~(PDE)
$$
\frac{\partial n}{\partial t} = \nabla \cdot (D(x,n) \nabla n) - \sigma n^2,
$$
where $n = n(x,t)$, $x\in \R^m$ and $t \ge 0$. Although the diffusion coefficient~$D(x,n)$ could depend on both~$x$ and $n$ in principle, as a first step and noting the observed weak diffusion of seismic activity~\citep{HeOuSo2003}, we will assume that it is a constant and study the nonlinear PDE
\begin{equation}
\label{rd-eq}
\frac{\partial n}{\partial t} = D \Delta n - \sigma n^2.
\end{equation}
For convenience, we shall refer to \eqref{rd-eq} as the `diffusive Omori law'.

The following questions will be addressed in this paper:
\begin{enumerate}
\setlength{\itemsep}{-0.5mm}
\item[Q1.] What is a spatio-temporal analogue of the Omori law~\eqref{omori}?
\item[Q2.] For a given initial number of active faults in the epicentral region of the main shock, say $n(x,0) = n_0(x)$, can we describe the evolution of the solution of \eqref{rd-eq} at later times?
\end{enumerate}

We will show that there is a similarity reduction of \eqref{rd-eq} that reduces the PDE to an ordinary differential equation~(ODE). Since constant solutions of the ODE lead to the Omori law, a nonconstant solution of the ODE can be viewed as a spatio-temporal analogue of the Omori law.

As \eqref{rd-eq} is a nonlinear PDE, it is not possible to find the general solution of the associated initial value problem~(IVP). Therefore the idea we will pursue here is to find bounding functions~$n^-$ and $n^+$ such that
$$
n^-(x,t) \le n(x,t) \le n^+(x,t), \quad x \in \R^m, \quad t \ge 0
$$
and for which analytical expressions for $n^\pm$ can be obtained. Of course, we have to find $n^\pm$ that are `close' in some sense for all~$x \in \R^m$ and $t \ge 0$. The method for constructing the upper and lower solutions for a family of reaction-diffusion equations is adapted from that in \citet{RoMi2002} and \citet{Ro2003}, which in turn has been extended recently in \citet{Ro2021}.

The outline of this paper is as follows. In Section~2, we propose a PDE model for aftershocks and exhibit a similarity reduction that reduces the PDE to an ODE. We show why a nonconstant solution of the ODE leads to a spatio-temporal analogue of the Omori-Utsu law. In Section~3, we consider the particular case when $p = 1$ and find an exact and explicit similarity solution of the `diffusive Omori law'. Section~4 investigates the spatio-temporal dynamics of the solution of the IVP for a generalised version of \eqref{rd-eq} with the help of a careful contruction of bounding functions. These bounding functions satisfy nonlinear PDEs similar to \eqref{rd-eq} but are linearisable to the diffusion equation. Brief concluding remarks are given in Section~5.

\section{A PDE model of aftershocks and a similarity reduction}

It is easy to see from the compound model~\eqref{otsuka} that
\begin{align*}
\frac{\d n}{\d t} & = -\alpha K \e^{-\alpha t} (t + c)^{-p} - p K \e^{-\alpha t} (t + c)^{-p - 1} = -\alpha n - \frac{p}{(K \e^{-\alpha t})^\frac{1}{p}} n^{1 + \frac{1}{p}},
\end{align*}
where $n = n(t)$. Now suppose that $n = n(x,t)$, where $x \in \R^m$ and $t \ge 0$. This naturally leads us to the consideration of the PDE
\begin{equation}
\label{n-pde}
\frac{\partial n}{\partial t} = D \Delta n -\alpha n - \frac{p}{(K \e^{-\alpha t})^\frac{1}{p}} n^{1 + \frac{1}{p}}.
\end{equation}
Note that setting $\alpha = 0$ and $p = 1$ in \eqref{n-pde} yields \eqref{rd-eq} with $\sigma = \frac{1}{K}$. Introducing the transformation~$n(x,t) = \e^{-\alpha t} u(x,t)$ in \eqref{n-pde}, we obtain the simpler PDE
\begin{equation}
\label{u-pde}
\frac{\partial u}{\partial t} = D \Delta u - \frac{p}{K^\frac{1}{p}} u^{1 + \frac{1}{p}}.
\end{equation}

Let us now look for a similarity reduction of \eqref{u-pde}. Suppose that
\begin{equation}
\label{ansatz}
u(x,t) = f(t) F(y), \quad y = g(t) \vert x \vert = g(t) (x_1^2 + \cdots + x_m^2)^\frac{1}{2},
\end{equation}
where $f$ ang $g$ are functions of $t$ to be determined such that \eqref{u-pde} becomes an ODE in $F = F(y)$. Straightforward calculations give
$$
\frac{\partial y}{\partial x_i} = g(t) \frac{x_i}{\vert x \vert}, \quad \frac{\partial^2 y}{\partial x_i^2} = -g(t) \frac{x_i^2}{\vert x \vert^3} + g(t) \frac{1}{\vert x \vert}, \quad \Big(\frac{\partial y}{\partial x_i}\Big)^2 = g(t)^2 \frac{x_i^2}{\vert x \vert^2},
$$
so that
$$
\Delta y = \sum_{i = 1}^m \frac{\partial^2 y}{\partial x_i^2} = (m - 1) g(t) \frac{1}{\vert x \vert}, \quad \vert \nabla y \vert^2 = \sum_{i = 1}^m \Big(\frac{\partial y}{\partial x_i}\Big)^2 = g(t)^2.
$$

Taking partial derivatives of \eqref{ansatz}, we have
$$
\frac{\partial u}{\partial t}(x,t) = f'(t) F(y) + f(t) g'(t) \vert x \vert F'(y),
$$
$$
\frac{\partial u}{\partial x_i} = f(t) F'(y) \frac{\partial y}{\partial x_i}, \quad \frac{\partial^2 u}{\partial x_i^2} = f(t) F'(y) \frac{\partial^2 y}{\partial x_i^2} + f(t) F''(y) \Big(\frac{\partial y}{\partial x_i}\Big)^2.
$$
Hence, substituting 
\begin{align*}
\Delta u = f(t) F'(y) \Delta y + f(t) F''(y) \vert \nabla y \vert^2 = (m - 1) f(t) g(t) \frac{1}{\vert x \vert} F'(y) + f(t) g(t)^2 F''(y)
\end{align*}
into \eqref{u-pde}, we obtain
\begin{align*}
& D (m - 1) f(t) g(t) \frac{1}{\vert x \vert} F'(y) + D f(t) g(t)^2 F''(y) - f'(t) F(y) - f(t) g'(t) \vert x \vert F'(y) \\
& \qquad {} - \frac{p}{K^\frac{1}{p}} f(t)^{1 + \frac{1}{p}} F(y)^{1 + \frac{1}{p}} = 0.
\end{align*}
Multiplying both sides by $\frac{1}{f(t) g(t)^2}$ and recalling \eqref{ansatz}, we get
\begin{equation}
\label{F-ode-temp}
D \Big[F''(y) + (m - 1) \frac{1}{y} F'(y)\Big] - \frac{g'(t)}{g(t)^3} y F'(y) - \frac{f'(t)}{f(t) g(t)^2} F(y) - \frac{p}{K^\frac{1}{p}} \frac{f(t)^\frac{1}{p}}{g(t)^2} F(y)^{1 + \frac{1}{p}} = 0.
\end{equation}

We require that the coefficients in \eqref{F-ode-temp} be independent of $t$, i.e.
\begin{equation}
\label{a-b-cond}
\frac{g'(t)}{g(t)^3} = a, \quad \frac{f'(t)}{f(t) g(t)^2} = b, \quad \frac{\d}{\d t}\Big[\frac{f(t)^\frac{1}{p}}{g(t)^2}\Big] = 0,
\end{equation}
where $a$ and $b$ are constants to be determined. The third equation in \eqref{a-b-cond} implies that
$$
g(t)^2 \frac{1}{p} f(t)^{\frac{1}{p} - 1} f'(t) = f(t)^\frac{1}{p} 2 g(t) g'(t).
$$
Dividing through by $f(t)^\frac{1}{p} g(t)^2$ and using \eqref{a-b-cond},
$$
\frac{1}{p} \frac{f'(t)}{f(t)} = 2 \frac{g'(t)}{g(t)} \quad \text{or} \quad \frac{b}{p} g(t)^2 = 2 a g(t)^2.
$$
Hence we must have $b = 2 a p$. It is not difficult to verify that
\begin{equation}
\label{g}
g(t) = (-2 a t + c_1)^{-\frac{1}{2}},
\end{equation}
where $c_1$ is an arbitrary constant of integration, satisfies the the first equation in \eqref{a-b-cond}. Combining the first and second equations in \eqref{a-b-cond}, we have
$$
\frac{f'(t)}{f(t)} = b g(t)^2 = \frac{b}{a} \frac{g'(t)}{g(t)} = 2 p \frac{g'(t)}{g(t)}, 
$$
which upon integrating yields
$$
\log(f(t)) = 2 p \log(g(t)) + \log(c_2),
$$
where $c_2$ is a (positive) arbitrary constant of integration. Then
\begin{equation}
\label{f}
f(t) = c_2 g(t)^{2 p} = c_2 (-2 a t + c_1)^{-p},
\end{equation}
where we substituted the expression for $g(t)$ from \eqref{g}. We see from \eqref{f} that
$$
\frac{f(t)^\frac{1}{p}}{g(t)^2} = c_2^\frac{1}{p}
$$
and therefore the third equation in \eqref{a-b-cond} holds. 

Let $c_1 = c \ge 0$, $c_2 = 1$ and $a = -\frac{1}{2}$, implying that $b = -p$. Thus \eqref{F-ode-temp} simplifies to the ODE
\begin{equation}
\label{F-ode}
D \Big[F''(y) + (m - 1) \frac{1}{y} F'(y)\Big] + \frac{1}{2} y F'(y) + p F(y) - \frac{p}{K^\frac{1}{p}} F(y)^{1 + \frac{1}{p}} = 0
\end{equation}
and a similarity reduction of the PDE~\eqref{u-pde} is given by
$$
u(x,t) = \frac{1}{(t + c)^p} F\Big(\frac{\vert x \vert}{\sqrt{t + c}}\Big). 
$$
Finally, a similarity reduction of the PDE~\eqref{n-pde} using \eqref{ansatz} is therefore
\begin{equation}
\label{n-pde-sol}
n(x,t) = \frac{\e^{-\alpha t}}{(t + c)^p} F\Big(\frac{\vert x \vert}{\sqrt{t + c}}\Big),
\end{equation}
where $F$ satisfies the ODE~\eqref{F-ode}.

\begin{rem}
In particular, suppose that $F$ is a nonzero constant solution of \eqref{F-ode}. Then $F(y) = K$ necessarily and \eqref{n-pde-sol} simplifies to 
$$
n(x,t) = \frac{K \e^{-\alpha t}}{(t + c)^p},
$$ 
which is \eqref{otsuka}. If $\alpha = 0$, this is of course the Omori-Utsu law~\eqref{utsu}. Hence, if $\alpha = 0$ and $F$ this time is any nonconstant solution of the ODE~\eqref{F-ode}, then we can think of \eqref{n-pde-sol} as  a spatio-temporal analogue of the Omori-Utsu law. 
\end{rem}

\section{Spatio-temporal Omori law:~an exact solution}

In this section, we fix $p = 1$ as in the Omori law~\eqref{omori}. Then \eqref{F-ode} becomes
\begin{equation}
\label{F-ode-omori}
D \Big[F''(y) + (m - 1) \frac{1}{y} F'(y)\Big] + \frac{1}{2} y F'(y) + F(y) - \frac{1}{K} F(y)^2 = 0.
\end{equation}
As \eqref{F-ode-omori} is a second-order ODE, two boundary conditions are required, e.g.
\begin{equation}
\label{F-ode-bc}
F(0) \text{ given}, \quad F(\infty) = 0.
\end{equation}

To find an exact solution of \eqref{F-ode-omori}, it is convenient to introduce a further change of variables. Let $z = y^2$ and $G(z) = F(y)^{-1}$, so that
\begin{equation}
\label{F-G-der}
F'(y) = - 2 y G(z)^{-2} G'(z), \quad \frac{1}{y} F'(y) = -2 G(z)^{-2} G'(z), \quad y F'(y) = -2 z G(z)^{-2} G'(z).
\end{equation}
It follows from the last equation in \eqref{F-G-der} that
$$
y F''(y) + F'(y) = 2 y \frac{\d}{\d z}[-2 z G(z)^{-2} G'(z)] 
$$
or, with the help of the second equation in \eqref{F-G-der},
\begin{equation}
\label{F-der2}
F''(y) = 2 \frac{\d}{\d z}[-2 z G(z)^{-2} G'(z)] + 2 G(z)^{-2} G'(z).
\end{equation}

Substituting \eqref{F-G-der} and \eqref{F-der2} into \eqref{F-ode-omori}, we see that
\begin{equation}
\label{G-ode1}
\begin{split}
& D\Big\{2 \frac{\d}{\d z}[-2 z G(z)^{-2} G'(z)] + 2 G(z)^{-2} G'(z) - 2 (m - 1) G(z)^{-2} G'(z)\Big\} \\
& \qquad {} - z G(z)^{-2} G'(z) + G(z)^{-1} - \frac{1}{K} G(z)^{-2} = 0.
\end{split}
\end{equation}
However, since
$$
\frac{\d}{\d z}[-2 z G(z)^{-2} G'(z)] = - 2  G(z)^{-2} G'(z) + 4 z G(z)^{-3} G'(z)^2 -2 z G(z)^{-2} G''(z),
$$
we get from \eqref{G-ode1} that
\begin{equation}
\label{G-ode2}
\begin{split}
& D[8 z G'(z)^2 - 4 z G(z) G''(z) - 2 m G(z) G'(z)] - z G(z) G'(z) + G(z)^2 - \frac{1}{K} G(z) = 0.
\end{split}
\end{equation}
The boundary conditions in \eqref{F-ode-bc} become
\begin{equation}
\label{G-ode-bc}
G(0) \text{ given}, \quad G(\infty) = \infty.
\end{equation}

After some trial and error, it turns out that we can find a rational function solution of \eqref{G-ode2} in the form of
\begin{equation}
\label{G-ansatz}
G(z) = \frac{k_1 + k_2 z + k_3 z^2}{k_4 + k_5 z},
\end{equation}
where $k_1$, $k_2$, $k_3$, $k_4$ and $k_5$ are constants to be determined. Without loss of generality, we may take $k_3 = 1$. Substituting \eqref{G-ansatz} into \eqref{G-ode2}, combining terms with like powers in $z$ and equating the corresponding coefficients of the powers of $z$ to zero, an algebraic system for the constants is generated. The solution consists of an elementary, but cumbersome, computation that is most conveniently checked by means of a Maxima routine. Three sets of solutions are obtained, namely
\begin{equation}
\label{k-sol1}
\begin{split}
k_1 & = \frac{D^2 [\pm (96 m^2 + 2528 m + 6976) + (4 m^2 + 472 m + 3424) \sqrt{2 m + 4}]}{\sqrt{2 m + 4} \pm 4}, \\
k_2 & = \frac{D[\pm (56 m + 304) + (4 m + 136) \sqrt{2 m + 4}]}{\sqrt{2 m + 4} \pm 4}, \\
k_4 & = D^2 K [288 m + 1152 \pm (24 m + 576) \sqrt{2 m + 4}], \\
k_5 & = D K (48 \pm 12 \sqrt{2 m + 4})
\end{split}
\end{equation}
and
\begin{equation}
\label{k-sol2}
k_1 = 0, \quad k_2 = \frac{D}{3} (2 m - 32), \quad k_4 = -\frac{D^2 K}{3} (4 m^2 - 80 m + 256), \quad k_5 = -D K (2 m - 8).
\end{equation}

Thus, from \eqref{n-pde-sol} an exact similarity solution of the PDE~\eqref{n-pde} when $p = 1$ is given by
\begin{equation}
\label{n-pde-sol-omori}
n(x,t) = \frac{\e^{-\alpha t}}{t + c} F\Big(\frac{\vert x \vert}{\sqrt{t + c}}\Big) = \frac{\e^{-\alpha t}}{t + c} \frac{1}{G(\frac{\vert x \vert^2}{t + c})}, \quad G(z) = \frac{k_1 + k_2 z + z^2}{k_4 + k_5 z},
\end{equation}
where $k_1$, $k_2$, $k_4$ and $k_5$ are as in \eqref{k-sol1} or \eqref{k-sol2}. Choosing the constants with the plus signs in \eqref{k-sol1} ensures that $n(x,t) \ge 0$ in \eqref{n-pde-sol-omori}. For each fixed $x \in \R^m$, we observe that
$$
\lim_{t \rightarrow \infty} n(x,t) = 0.
$$
since $\alpha \ge 0$ and $G(0) = \frac{k_1}{k_4} \ne 0$. In the special case when $\alpha = 0$ and $D = K = 1$, \eqref{n-pde-sol-omori} recovers the exact similarity solution found by \citet{HeRo2005} for the diffusive Smoluchowski's equations. A plot of $n(x,0.5)$ from \eqref{n-pde-sol-omori}, with $\alpha = 0.01$, $c = 0$, $D = 0.01$, $K = 5$ and $m = 1$, is shown in Figure~1.
\begin{center}
\begin{figure}[h]
\centering
\includegraphics[width=.8\linewidth]{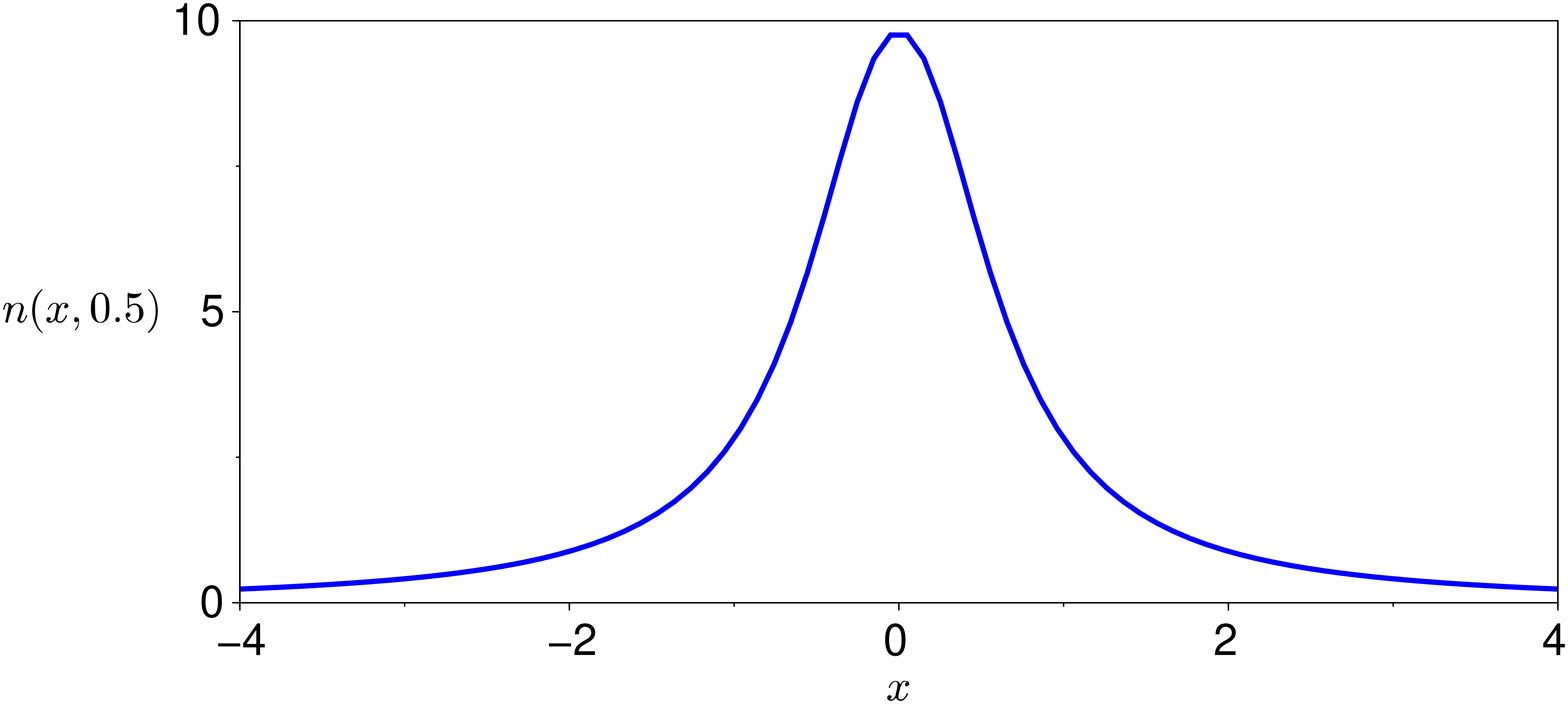}
\caption{Plot of $n(x,0.5)$ from \eqref{n-pde-sol-omori}, where $\alpha = 0.01$, $c = 0$, $D = 0.01$, $K = 5$ and $m = 1$.}
\end{figure}
\end{center}

\begin{rem}
The results in this section will not necessarily hold for $p \ne 1$ (i.e.~the non-Omori case) since the ansatz~\eqref{G-ansatz} may not be valid for an arbitrary~$p > 0$. 
\end{rem}

\section{Bounds for the solution of an IVP for a diffusive Omori-Utsu law}

In this section, we consider the IVP
\begin{equation}
\label{ivp-omori-n}
\begin{split}
& \frac{\partial n}{\partial t} = D \Delta n -\alpha n - \frac{p}{(K \e^{-\alpha t})^\frac{1}{p}} n^{1 + \frac{1}{p}}, \quad x \in \R^m, \quad t > 0, \\
& n(x,0) = n_0(x), \quad x \in \R^m.
\end{split}
\end{equation}
For the initial distribution, we assume that 
\begin{equation}
\label{init-ass}
\mu \le n_0(x) \le K, \quad x \in \R^m,
\end{equation}
where $0 < \mu < K$. 

If $n(x,t) = \e^{-\alpha t} u(x,t)$, then \eqref{ivp-omori-n} is equivalent to the IVP 
\begin{equation}
\label{ivp-omori-u}
\begin{split}
& \frac{\partial u}{\partial t} = D \Delta u - \frac{p}{K^\frac{1}{p}} u^{1 + \frac{1}{p}}, \quad x \in \R^m, \quad t > 0, \\
& u(x,0) = n_0(x), \quad x \in \R^m.
\end{split}
\end{equation}

Although the class of PDEs considered in \cite{RoMi2002}, \cite{Ro2003} and \cite{Ro2021} was of Fisher-KPP type, the procedure for constructing upper and lower solutions for \eqref{ivp-omori-u} remains applicable here. Indeed, let $u^\pm = u^\pm(x,t)$ satisfy the initial value problems
\begin{equation}
\label{u-upper-lower-pdes}
\begin{split}
& \frac{\partial u^\pm}{\partial t} = D \Delta u^\pm - \frac{p}{K^\frac{1}{p}} (u^\pm)^{1 + \frac{1}{p}} - D \Big(1 + \frac{1}{p}\Big) \frac{(u^\pm)^\frac{1}{p} \mp K^\frac{1}{p}}{(u^\pm)^{1 + \frac{1}{p}}} \vert \nabla u^\pm \vert^2, \quad x \in \R^m, \quad t > 0,\\
& u^\pm(x,0) = n_0(x), \quad x \in \R^m.
\end{split}
\end{equation}
Define the operator
$$
N(u) = D \Delta u - \frac{\partial u}{\partial t} - \frac{p}{K^\frac{1}{p}} u^{1 + \frac{1}{p}}.
$$
Because of the assumption~\eqref{init-ass}, we see that $N(u^+) \le 0$ and $N(u^-) \ge 0$. Hence $u^+$ is an upper solution while $u^-$ is a lower solution, i.e.
\begin{equation}
\label{bounds}
u^-(x,t) \le u(x,t) \le u^+(x,t), \quad x \in \R^m, \quad t \ge 0.
\end{equation}

It is not difficult to show that the nonlinear IVPs in \eqref{u-upper-lower-pdes} can be linearised through the transformations
\begin{equation}
\label{u-upper-lower-sol}
u^\pm(x,t) = \frac{K}{[1 \mp \frac{1}{p + 1} \log(v^\pm(x,t))]^p}
\end{equation}
to obtain the linear IVPs
\begin{equation}
\label{v-upper-lower-pdes}
\begin{split}
& \frac{\partial v^\pm}{\partial t} = D \Delta v^\pm \mp (p + 1) v^\pm, \quad x \in \R^m, \quad t > 0, \\
& v^\pm(x,0) = \e^{\mp (p + 1) [(\frac{K}{n_0(x)})^\frac{1}{p} - 1]}, \quad x \in \R^m.
\end{split}
\end{equation}
In turn, \eqref{v-upper-lower-pdes} can be mapped to corresponding IVPs for the diffusion equation, which can be solved explicitly. Therefore
\begin{equation}
\label{v-upper-lower-sol}
v^\pm(x,t) = \frac{1}{(4 \pi D t)^\frac{m}{2}} \int_{\R^m} \exp\Big(\mp (p + 1) t - \frac{\vert x - \xi \vert^2}{4 D t} \mp (p + 1) \Big[\Big(\frac{K}{n_0(\xi)}\Big)^\frac{1}{p} - 1\Big]\Big) \, \d \xi. 
\end{equation}
Recalling that $n(x,t) = \e^{-\alpha t} u(x,t)$ and using \eqref{u-upper-lower-sol} in \eqref{bounds}, we deduce that the solution of \eqref{ivp-omori-n} is bounded by
\begin{equation}
\label{n-upper-lower}
\frac{K \e^{-\alpha t}}{[1 + \frac{1}{p + 1} \log(v^-(x,t))]^p} \le n(x,t) \le \frac{K \e^{-\alpha t}}{[1 - \frac{1}{p + 1} \log(v^+(x,t))]^p}, \quad x \in \R^m, \quad t \ge 0,
\end{equation}
where $v^\pm$ are as given in \eqref{v-upper-lower-sol}. Equation~\eqref{n-upper-lower} describes the spatio-temporal dynamics of the `diffusive compound Omori-Utsu law'~\eqref{n-pde}. 

Next, we show some numerical results. Suppose that $D = 0.001$, $K = 5$, $m = 1$, $\alpha = 0$ and $p = 1.2$. To solve the IVP~\eqref{ivp-omori-n} numerically, we consider the finite domain~$-L \le x \le L$, where $L = 20$, and assume homogeneous Neumann boundary conditions. For a localised initial condition, the chosen boundary conditions will not affect the solution of \eqref{ivp-omori-n} substantially. For the initial distribution let us take
$$
n_0(x) = 0.05 + 0.8 \e^{-\vert x + \frac{L}{2}\vert} + 0.4 \e^{-\vert x - \frac{L}{2}\vert}.
$$
We can verify that \eqref{init-ass} is satisfied, with $\mu = 0.05$. In Figure~2, we plot the time evolution of the solution of \eqref{ivp-omori-n} (green curve) and the bounds in \eqref{n-upper-lower} (red and blue curves), where $v^\pm$ are described by \eqref{v-upper-lower-sol}.
\begin{figure}
\begin{subfigure}{0.5\textwidth}
  \centering
  \includegraphics[width=1.0\linewidth]{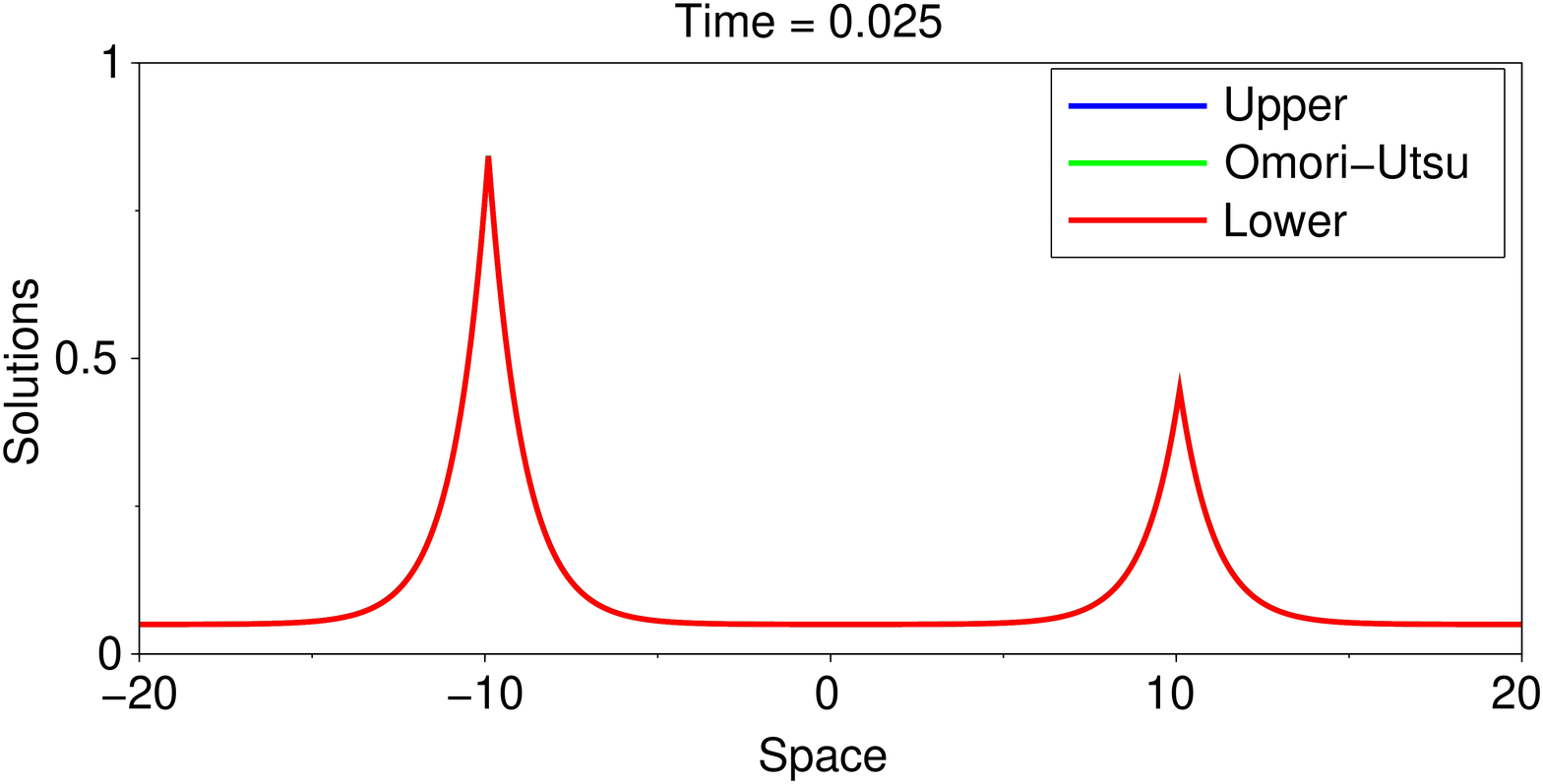}  
\end{subfigure}
\begin{subfigure}{.5\textwidth}
  \centering
  \includegraphics[width=1.0\linewidth]{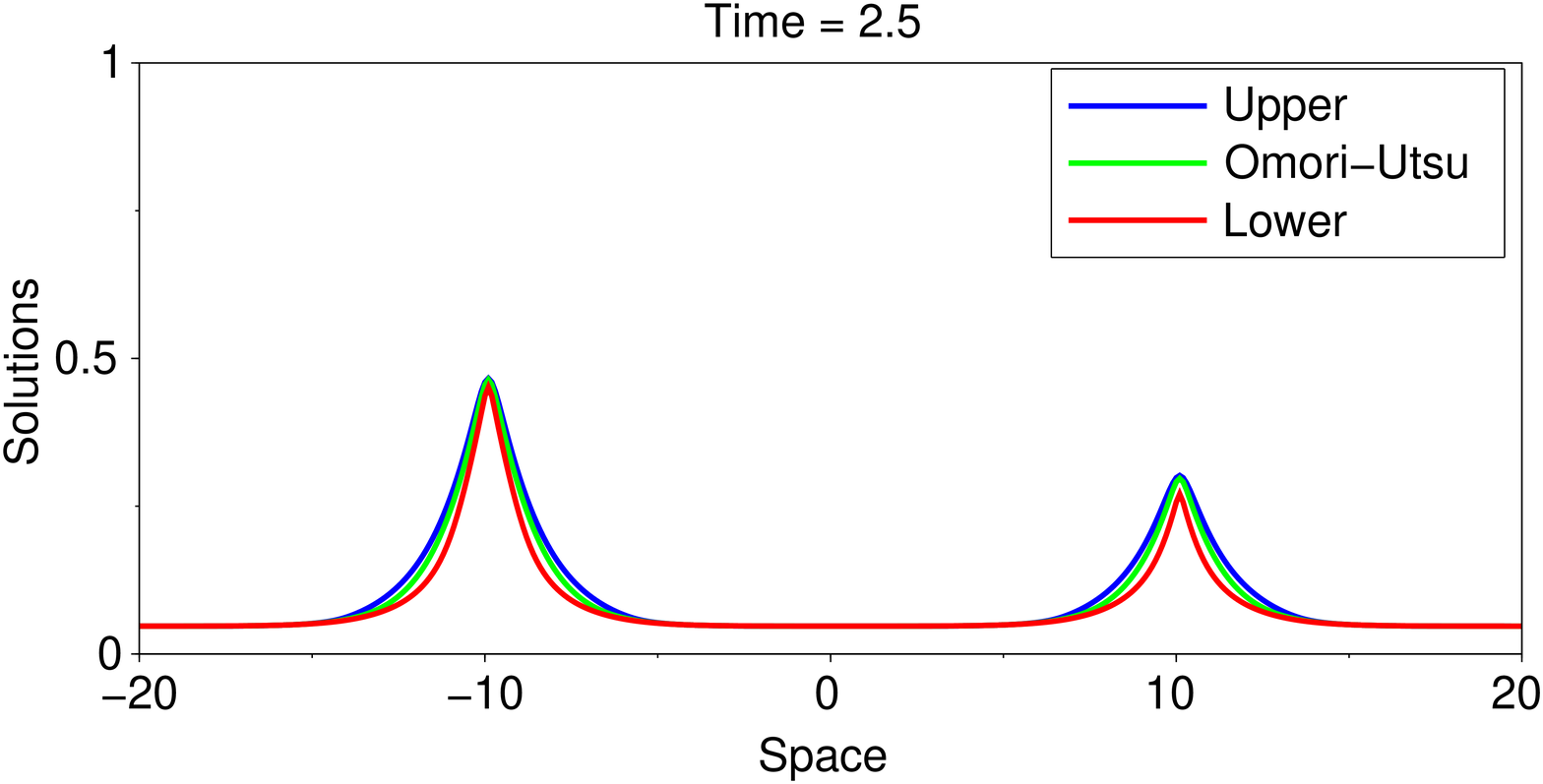}  
\end{subfigure}
\newline
\begin{subfigure}{.5\textwidth}
  \centering
  \includegraphics[width=1.0\linewidth]{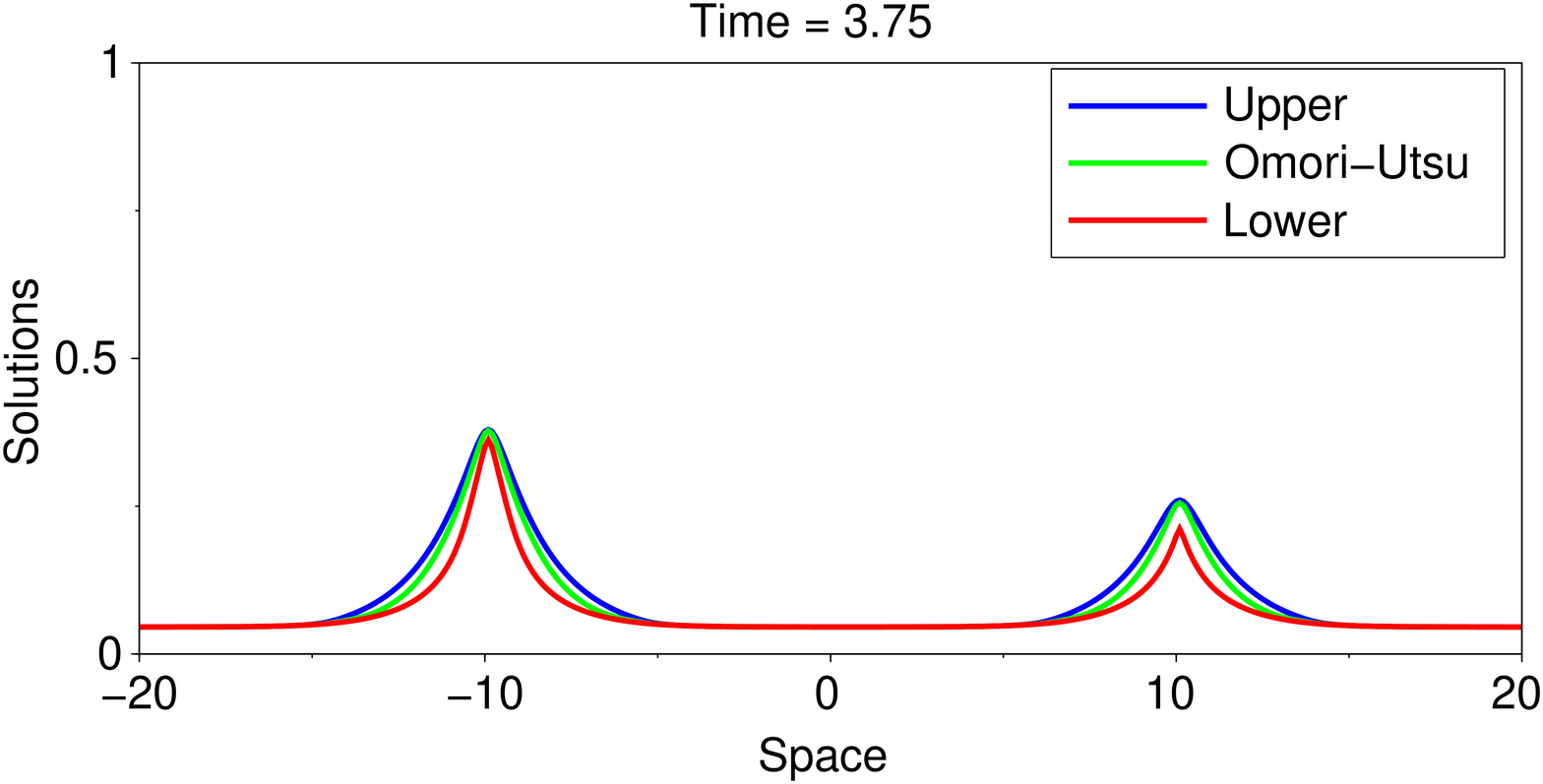}  
\end{subfigure}
\begin{subfigure}{.5\textwidth}
  \centering
  \includegraphics[width=1.0\linewidth]{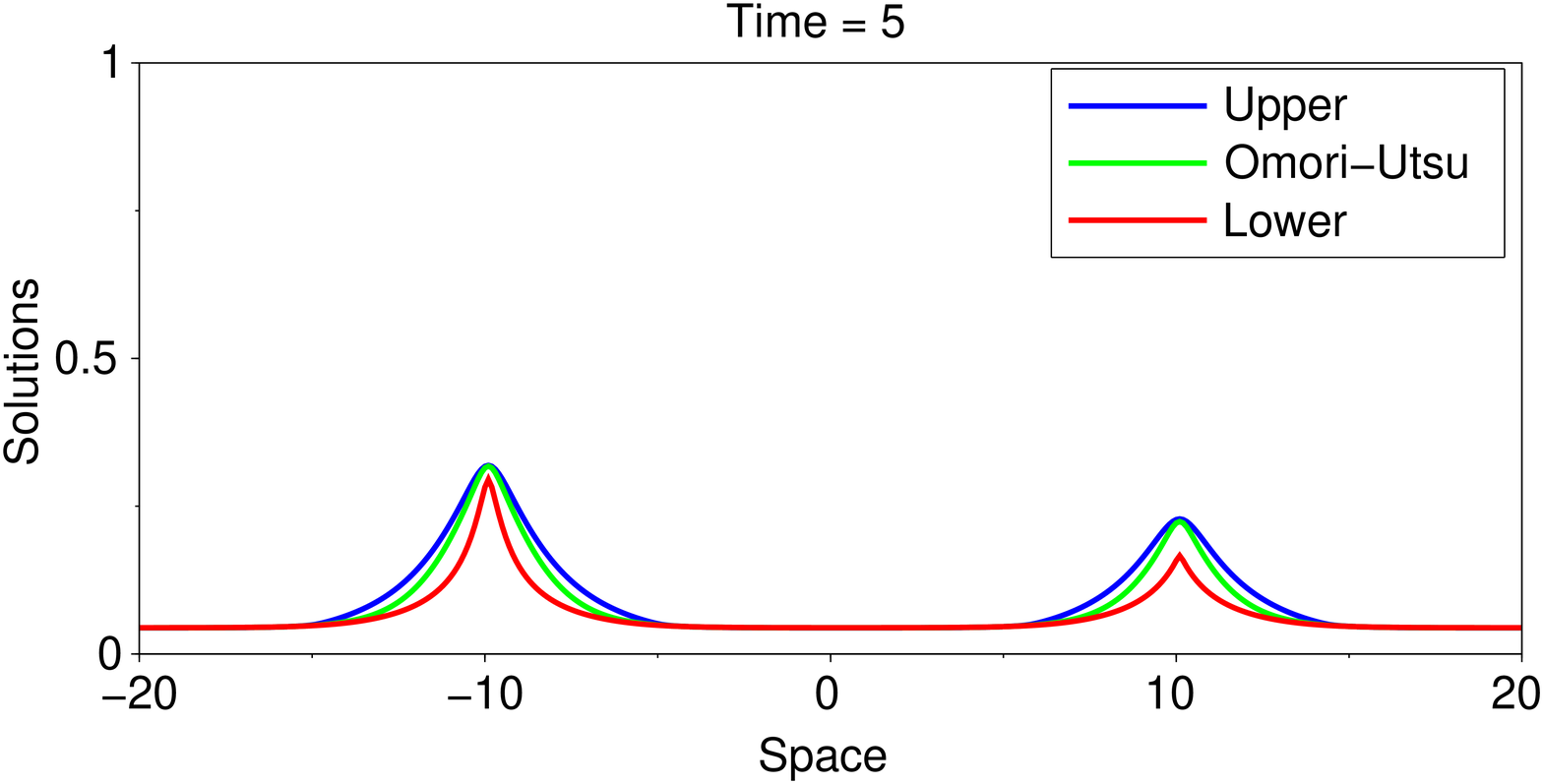}  
\end{subfigure}
\caption{Time snapshots of $n(x,t)$ (from \eqref{ivp-omori-n}, in green) and $\e^{-\alpha t} u^\pm(x,t)$ (from \eqref{n-upper-lower} and \eqref{v-upper-lower-sol}, in blue and red) for $t = 0.025, 2.5, 3.75, 5$, where $D = 0.001$, $K = 5$, $m = 1$, $\alpha = 0$ and $p = 1.2$.}
\label{fig:fig}
\end{figure}

\begin{rem}
We observe from \eqref{init-ass} that
$$
\exp\Big(-(p + 1) \Big[\Big(\frac{K}{n_0(\xi)}\Big)^\frac{1}{p} - 1\Big]\Big) \le 1
$$
for all $\xi \in \R^m$. Hence \eqref{v-upper-lower-sol} implies that
$$
\vert v^+(x,t) \vert \le \frac{1}{(4 \pi D t)^\frac{m}{2}} \int_{\R^m} \exp\Big(-(p + 1) t {-\frac{\vert x - \xi \vert^2}{4 D t}}\Big) \, \d \xi \le \e^{-(p + 1) t}
$$
and so $v^+(x,t) \rightarrow 0$ as $t \rightarrow \infty$, for every $x \in \R^m$. Similarly, \eqref{v-upper-lower-sol} shows that $v^-(x,t) \rightarrow \infty$ as $t \rightarrow \infty$, for every $x \in \R^m$. Therefore we deduce from \eqref{n-upper-lower} that 
$$
\lim_{t \rightarrow \infty} n(x,t) = 0
$$
for each fixed~$x \in \R^m$. This behaviour is analogous to that of the (spatially homogeneous) Omori-Utsu law~\eqref{utsu}.
\end{rem}

\section{Concluding remarks}

As mentioned in the Introduction, there are several ways of interpreting aftershock diffusion. In this article, we followed the interpretation put forward by \citet{Gu2017} for the temporal evolution and incorporated a diffusion term. We determined a spatio-temporal analogue of the Omori-Utsu law from a similarity reduction of the PDE for any $p > 0$. When $p = 1$, an exact and explicit similarity solution was found. It is not clear whether analytical solutions can be found for $p \ne 1$.  

We also considered an IVP for the `diffusive Omori-Utsu law'~\eqref{n-pde} with quite general initial distributions. Using appropriate upper and lower solutions, we were able to describe the spatio-temporal dynamics of the solution starting from a given initial number of active faults in the epicentral region of the main shock. As the numerical results in Figure~2 suggest, the two peaks in the initial distribution could represent the sites where the aftershocks are located close to the rupture plane of the main shock and the decay both in space and time can be observed. It remains to be seen whether the inclusion of diffusion in the Omori-Utsu law is in fact a satisfactory intepretation of aftershock diffusion.

\end{document}